\documentclass[iop]{emulateapj}
\usepackage{hyperref}
\usepackage{ulem}

\shorttitle{Magnetars radio emission}
\shortauthors{Szary et al.}

\begin{document}

    \title{On the Origin of Radio Emission from Magnetars}

    \author{Andrzej Szary\altaffilmark{1}, George I. Melikidze\altaffilmark{1,2}, Janusz Gil\altaffilmark{1}}
    
    \email{aszary@astro.ia.uz.zgora.pl}

    \altaffiltext{1}{Kepler Institute of Astronomy, University of Zielona G\'ora,\\ Lubuska 2, 65-265 Zielona G\'ora, Poland}
    \altaffiltext{2}{Abastumani Astrophysical Observatory, Ilia State University, 3-5 Cholokashvili Avenue, Tbilisi, 0160, Georgia}
 
  \begin{abstract}
    Magnetars are the most magnetized objects in the known universe.
    Powered by the magnetic energy, and not by the rotational energy as in the case of radio pulsars, they have long been regarded as a completely different class of neutron stars.
    The discovery of pulsed radio emission from a few magnetars weakened the idea of a clean separation between magnetars and normal pulsars.
    We use the partially screened gap (PSG) model to explain radio emission of magnetars.
    The PSG model requires that the temperature of the polar cap is equal to the so-called critical value, i.e. the temperature at which the thermal ions outflowing from the stellar surface screen the acceleration gap.
    We show that a magnetar has to fulfill the temperature, power and visibility conditions in order to emit radio waves.
    Firstly, in order to form PSG, the residual temperature of the surface has to be lower than the critical value. 
    Secondly, since the radio emission is powered by the rotational energy, it has to be high enough to enable heating of the polar cap by backstreaming particles to the critical temperature.
    Finally, the structure of the magnetic field has to be altered by magnetospheric currents in order to widen a radio beam and increase the probability of detection.
    Our approach allows us to predict whether a magnetar can emit radio waves using only its rotational period, period derivative, and surface temperature in the quiescent mode.
  \end{abstract}
  
    \keywords{pulsars: general --- stars: magnetars --- stars: neutron}
  
  \section{Introduction}
  
    Historically, magnetars were divided into two distinct classes, anomalous X-ray pulsars (AXPs) and soft gamma repeaters (SGRs), based on the way they were discovered.
    However, nowadays it is widely accepted that those objects are highly magnetized neutron stars, which can explain the properties of both AXPs and SGRs \citep{1995_Thompson, 2008_Mereghetti}.
    Unlike other neutron stars for which emission is powered by rotational energy, accretion, or heat of a star, the high-energy emission of magnetars is powered by the energy of their strong magnetic field.
    Assuming a dipolar configuration of the magnetic field and that the observed spin-down is caused by the emission of a rotating dipole in vacuum, \footnote{Although the assumptions are unrealistic, more realistic models (e.g., \citealt{2006_Spitkovsky}) give values within a factor of two from the simplified formula.} we can estimate the field strength at the polar cap as $B_{\rm d} \approx 6.4 \times 10^{19} \sqrt{P \dot{P}}\,{\rm G}$, where $P$ is the rotational period in seconds, and $\dot{P}$ is the period derivative.
    Using this formula, we can estimate the magnetic field of magnetars at the stellar surface of the order of $\sim 10^{14}-10^{15}\,{\rm G}$.
    However, this argument by itself cannot be treated as the final evidence of a strong magnetic field, owing to, for example, an unknown contribution of other processes to the observed torque (e.g., magnetar wind: \cite{1999_Harding}).
    Nevertheless, analysis of spectral and energy properties of AXPs/SGRs further confirmed strong magnetic fields of magnetars \citep[see review paper by][]{2013_Mereghetti}. 
  
    In the past few years it has been realized that the separation between magnetars and pulsars is not as sharp as previously considered.
    It is believed that the radio emission of pulsars is powered by the rotational energy of the star (also called spin-down luminosity) $L_{\rm SD} = \dot{E} = 4 \pi^2 I \dot{P} P^{-3} \simeq 3.95\times 10^{31} I_{45}\left ( \dot{P}/10^{-15} \right ) \left ( P/s \right )^{-3} \,{\rm erg \, s^{-1}}$, where $I=10^{45} I_{45} \,{\rm g\, cm^2}$ is the star's moment of inertia. 
    As we will show in the accompanying paper \citep{2015_Szary}, radio emission of pulsars mainly depends on details of charged particle acceleration in close vicinity of the pulsar's polar cap (an altitude $\lesssim 10^4\,{\rm cm}$).
    X-ray observations of old radio pulsars show that the hot spot surface, and thereby the actual polar cap, is much smaller than would result from the purely dipolar geometry.
    Using the flux conservation law we can estimate the magnetic field strength in the acceleration region to be of the order of $\sim 10^{14} \, {\rm G}$.
    The open question is why most magnetars do not emit radio waves since plasma responsible for radio emission is accelerated in similar conditions to those in radio pulsars.  
    Based on the observed sample of radio magnetars, \cite{2011_Melikidze, 2012_Rea} suggested that magnetars can be radio active when the quiescent X-ray luminosity is smaller than the spin-down luminosity.
    However, as shown by the latest spectral fits by \cite{2013_Vigano}, it is not the case for \mbox{XTE J1810-197} (see Table \ref{tab:magnetars}).
    
    In this paper, we study whether the inner acceleration region of magnetars can be described by the same model as in normal radio pulsars, namely the partially screened gap (PSG) model.
    Since most magnetars are radio quiet, we are focusing on finding conditions that should be met to form the PSG, and hence to produce plasma responsible for radio emission.
    
    
\begin{table*}
    \caption{Observed properties of magnetars. The individual columns are as follows: (1) Number of the magnetar, (2) Magnetar name, (3) Rotational period, (4) Period derivative, (5) Magnetic field strength at the polar cap, (6) Spin-down luminosity, (7) Radius of the polar cap, (8) Critical temperature (see Equation \ref{t_crit}), (9) Observed temperature, (10) Power required to heat up the polar cap to the critical temperature (see Equation \ref{l_heat}), (11) Observed X-ray luminosity, (12) Reference. Magnetars are sorted by name (2). The names written in italic correspond to magnetars with detected radio emission.
     }
    \label{tab:magnetars}
    {\tiny
    \begin{tabular}{llcccccccccc}
        \hline
        & & & & & & & & & & &  \\
        No. &  Name    &   $P$   &   $\dot{P}$   &   $B_{\rm d}$   &  $\log L_{\rm SD}$  &   $R_{\rm pc}$   &   $T_{\rm crit}$   &   $T_{\rm bb}$   &  $\log L_{\rm heat}$ & $\log L_{\rm bb}$ & Ref.  \\
        &  &   {$\left ( {\rm s} \right )$}   &   {$\left ( 10^{-11}\,{\rm s\,s^{-1}} \right )$}   &   {$\left ( 10^{14}\,{\rm G} \right )$}   &  {$\left ( {\rm erg\,s^{-1}} \right )$}   &  {$\left ( {\rm m} \right )$} &  {$\left ( {\rm 10^{6} K} \right )$} &  {$\left ( {\rm 10^{6} K} \right )$} &  {$\left ( {\rm erg\,s^{-1}} \right )$} & {$\left ( {\rm erg\,s^{-1}} \right )$}  & \\
        & & & & & & & & & & & \\
        \hline
        \hline
 1  &  1E 1048.1-5937  &   $6.46$  &  $1.25 - 5.00$  &  $5.75 - 11.50$  &  $33.26 - 33.87$  &  59  &  $7.43 - 12.49$  &  $7.43$  &  $31.28 - 32.18$  & $33.80 - 34.50$ &  1, 2   \\ 
 2  &  {\it 1E 1547.0-5408}  &   $2.07$  &  $2.60 - 9.78$  &  $4.70 - 9.11$  &  $35.06 - 35.64$  &  104  &  $6.38 - 10.49$  &  $6.03$  &  $31.51 - 32.37$  & $34.30 - 34.70$ &  3, 4, 2  \\ 
 3  &  1E 1841-045  &   $11.79$  &  $4.09$  &  $14.06$  &  $32.99$  &  44  &  $14.52$  &  $5.57$  &  $32.18$  & $35.20 - 35.50$ &  5, 2   \\ 
 4  &  1E 2259+586  &   $6.98$  &  $0.05$  &  $1.18$  &  $31.75$  &  57  &  $2.26$  &  $4.64$  &  $29.17$  & $35.00 - 35.40$ &  5, 2   \\ 
 5  &  1RXS J170849.0-400910  &   $11.01$  &  $1.95$  &  $9.36$  &  $32.76$  &  45  &  $10.71$  &  $5.22$  &  $31.68$  & $34.80 - 35.10$ &  5, 2   \\ 
& & & & & & & & & & \\ 
 6  &  3XMM J185246.6+003317  &   $11.56$  &  $0.01$  &  $0.81$  &  $30.55$  &  44  &  $1.71$  &  --  &  $28.48$  & $30.78$ &  6   \\ 
 7  &  4U 0142+61  &   $8.69$  &  $0.20$  &  $2.68$  &  $32.09$  &  51  &  $4.19$  &  $4.76$  &  $30.15$  & $35.40 - 35.80$ &  5, 7   \\ 
 8  &  CXOU J010043.1-721134  &   $8.02$  &  $1.88$  &  $7.86$  &  $33.16$  &  53  &  $9.39$  &  $4.06$  &  $31.59$  & $35.20 - 35.50$ &  8, 2   \\ 
 9  &  CXOU J164710.2-455216  &   $10.61$  &  $0.04$  &  $1.32$  &  $31.12$  &  46  &  $2.46$  &  $3.83$  &  $29.14$  & $33.10 - 33.60$ &  9, 2   \\ 
 10  &  CXOU J171405.7-381031  &   $3.83$  &  $5.88 - 10.50$  &  $9.60 - 12.83$  &  $34.62 - 34.87$  &  77  &  $10.91 - 13.56$  &  $6.27$  &  $32.17 - 32.55$  & $34.90 - 35.20$ &  10, 2   \\ 
 11  &  {\it PSR J1622-4950}  &   $4.33$  &  $0.94 - 1.94$  &  $4.09 - 5.86$  &  $33.66 - 33.98$  &  72  &  $5.75 - 7.54$  &  $5.80$  &  $31.01 - 31.48$  & $32.64$ &  11, 12  \\ 
& & & & & & & & & & \\ 
 12  &  SGR 0418+5729  &   $9.08$  &  $0.004$  &  $0.12$  &  $29.32$  &  50  &  $0.41$  &  $3.71$  &  $26.11$  & $30.70 - 31.10$ &  13   \\ 
 13  &  SGR 0501+4516  &   $5.76$  &  $0.59$  &  $3.74$  &  $33.09$  &  62  &  $5.38$  &  $6.61$  &  $30.77$  & $33.20 - 34.00$ &  14, 2   \\ 
 14  &  SGR 0526-66  &   $8.05$  &  $3.80$  &  $11.20$  &  $33.46$  &  53  &  $12.24$  &  $5.57$  &  $32.05$  & $35.40 - 35.80$ &  15, 2   \\ 
 15  &  SGR 1627-41  &   $2.59$  &  $1.90$  &  $4.49$  &  $34.63$  &  93  &  $6.17$  &  $5.22$  &  $31.35$  & $34.40 - 34.80$ &  16, 17, 2   \\ 
 16  &  SGR 1806-20  &   $7.55$  &  $8.27 - 79.00$  &  $15.99 - 49.42$  &  $33.88 - 34.86$  &  55  &  $15.99 - 37.28$  &  $8.01$  &  $32.54 - 34.01$  & $35.10 - 35.50$ &  18, 2   \\ 
 17  &  SGR 1833-0832  &   $7.57$  &  $0.35$  &  $3.29$  &  $32.50$  &  55  &  $4.89$  &  --  &  $30.48$  & $33.00 - 35.00$ &  19, 20   \\ 
 18  &  SGR 1900+14  &   $5.20$  &  $6.13 - 20.00$  &  $11.42 - 20.64$  &  $34.24 - 34.75$  &  66  &  $12.43 - 19.37$  &  $4.53$  &  $32.26 - 33.04$  & $35.00 - 35.40$ &  21, 2   \\ 
 19  &  {\it SGR J1745-2900}  &   $3.76$  &  $0.61 - 1.39$  &  $3.07 - 4.62$  &  $33.66 - 34.01$  &  77  &  $4.64 - 6.30$  &  --  &  $30.69 - 31.23$  & $32.04$ &  22, 23  \\ 
 20  &  Swift J1822.3-1606  &   $8.44$  &  $0.002 - 0.01$  &  $0.27 - 0.54$  &  $30.15 - 30.74$  &  52  &  $0.75 - 1.25$  &  $6.27$  &  $27.18 - 28.07$  & $32.90 - 33.20$ &  24, 2   \\ 
 21  &  Swift J1834.9-0846  &   $2.48$  &  $0.80$  &  $2.84$  &  $34.31$  &  95  &  $4.38$  &  --  &  $30.77$  & $30.92$ &  25, 26   \\ 
 22  &  {\it XTE J1810-197}  &   $5.54$  &  $0.43 - 1.04$  &  $3.12 - 4.87$  &  $33.00 - 33.38$  &  64  &  $4.70 - 6.55$  &  $3.02$  &  $30.55 - 31.12$  & $34.00 - 34.40$ &  27, 28, 2  \\ 

      & & & & & & & & & & & \\
      \hline
      \hline
  \end{tabular}
  {\bf References.} (1) \citealt{2009_Dib}, (2) \citealt{2013_Vigano}, (3) \citealt{2012_Dib}, (4) \citealt{2011_Bernardini}, (5) \citealt{2014_Dib}, (6) \citealt{2014_Rea}, (7) \citealt{2007_Rea}, (8) \citealt{2005_McGarry}, (9) \citealt{2013_An}, (10) \citealt{2010_Sato}, (11) \citealt{2010_Levin}, (12) \citealt{2012_Anderson}, (13) \citealt{2013_Rea}, (14) \citealt{2014_Camero}, (15) \citealt{2009_Tiengo}, (16) \citealt{2009_Esposito}, (17) \citealt{2009_Esposito+b}, (18) \citealt{2007_Woods}, (19) \citealt{2011_Esposito}, (20) \citealt{2010_Gogus}, (21) \citealt{2006_Mereghetti}, (22) \citealt{2014_Kaspi}, (23) \citealt{2013_Mori}, (24) \citealt{2014_Scholz}, (25) \citealt{2012_Kargaltsev}, (26) \citealt{2012_Younes}, (27) \citealt{2007_Camilo}, (28) \citealt{2011_Bernardini+b}
  }
\end{table*}   
    
    \newpage
  \section{Partially Screened Gap} \label{sec:psg}
    The acceleration gap above the polar cap can form if a local charge density is lower than the corotational charge density \citep{1969_Goldreich}.
    The charge depletion in this region depends on the binding energy of the positive ${\rm _{26}^{56}Fe}$ ions in the crust.
    The binding energy, and thus emission of iron ions from the condensed stellar surface was calculated by \cite{2007_Medin}.
    The critical temperature, i.e. the temperature at which charge density of ions is equal to the corotational charge density, can be described as:
    \begin{equation}
        T_{{\rm crit}} \approx 2.0 \times10^{6} B_{14}^{0.75},
        \label{t_crit}
    \end{equation}
    where $B_{14} = B_{{\rm s}}/\left(10^{14}\,{\rm G}\right)$ is a surface magnetic field at the polar cap.
    Spectral fits to the X-ray data of old radio pulsars show that the temperature of polar caps is about a few million kelvin.
    Furthermore, the X-ray observations allow us to indirectly determine the surface magnetic field of radio pulsars using the magnetic flux conservation law $B_{\rm s} = B_{\rm d} A_{\rm pc} / A_{\rm bb}$, where $A_{\rm pc}\approx6.2\times10^{4}P^{-1}\,{\rm m^{2}}$ is the conventional polar cap area (assuming purely dipolar configuration of magnetic field) and $A_{\rm bb}$ is the observed polar cap area.
    Although still the subject of controversy as the X-ray observations are burdened with large uncertainties, the observed temperature and the magnetic field strength at the polar cap of radio pulsars agree with the theoretical predictions of the critical temperature \citep[see, e.g.,][and references therein]{2013_Szary}.
    In order to form the PSG and explain both radio and thermal X-ray emissions of normal pulsars, the surface magnetic field has to be dominated by very strong crust-anchored magnetic anomalies \citep{2002_Gil, 2013_Szary}; thus, $B_{\rm s} \sim 10^{14} \,{\rm G} \gg B_{\rm d}$.
	
    The basic features of the PSG model \citep{2003_Gil} are as follows.
    The supply rate of positive charges from the stellar surface is not enough to compensate the outflow of charges through the light cylinder.
    As a consequence, it leads to the development of a potential drop above the polar cap.
    The backstreaming electrons accelerated in the gap heat the polar cap.
    Depending on the mode, either the polar cap is overheated by electrons, leading to its breakdown (the PSG-off mode), or the temperature is kept close, but still below, the critical temperature (the PSG-on mode; \cite{2015_Szary}).
    In the PSG-on mode the surface temperature $T_{\rm s}$ is thermostatically regulated, leading to a continuous outflow of iron ions which leads to a partial screening of the acceleration potential drop.
    The gap breakdown in the PSG-on mode is due to the production of dense electron-positron plasma ($\rho_{\rm p} \gg \rho_{\rm GJ}$) which is responsible for the generation of radio emission at higher altitudes. 
    Regardless of the mode in which PSG operates, the observed temperature of a few million kelvin requires a magnetic field of the order of $10^{14}\,{\rm G}$ (see Equation \ref{t_crit}).
    Since the dipolar component of the magnetic field at the polar cap of normal radio pulsars is of the order of $10^{12} \, \rm G$, formation of the PSG requires a much stronger, and thus highly nondipolar magnetic field at the polar cap.
    Note that the nondipolar configuration of the surface magnetic field was proposed from the very beginning of pulsar astronomy, e.g., the vacuum gap  model \citep{1975_Ruderman} requires a highly nondipolar radius of curvature, $\Re \approx 10^{6} \,{\rm cm}$, in order to enable absorption of \mbox{$\gamma$-photons} in a gap region and electron-positron pair production.
    In the case of magnetars, on the other hand, the dipolar component of the magnetic field at the surface already fulfills the PSG model requirement of strong magnetic field $\sim 10^{14}\,{\rm G}$.
    Furthermore, in one of the modes of PSG (namely the PSG-on mode) inverse Compton scattering is responsible for $\gamma$-photon emission in a gap region.
    Even assuming dipolar curvature of magnetic field lines, owing to the high energy of such $\gamma$-photons they are easily converted to electron-positron pairs. 
    Thus, unlike normal pulsars, for magnetars it is not required to have a nondipolar configuration of the surface magnetic field to form a PSG.
    \cite{2013_Geppert, 2014_Geppert} showed that the Hall drift is the physical process that can be responsible for production of small-scale strong surface magnetic field anomalies on timescales of $10^4\,{\rm yr}$.
    The timescale suggest that magnetars, as young neutron stars, should be characterized by a dipolar surface magnetic field.
    The surface magnetic field of normal radio pulsars, on the other hand, should be dominated by magnetic spots produced by means of nonlinear interaction between poloidal and toroidal components of the subsurface magnetic field.
    

  \section{Results}

    \subsection{Temperature} \label{sec:temperature}
    
    \begin{figure}[t]
        \begin{center}
            \includegraphics{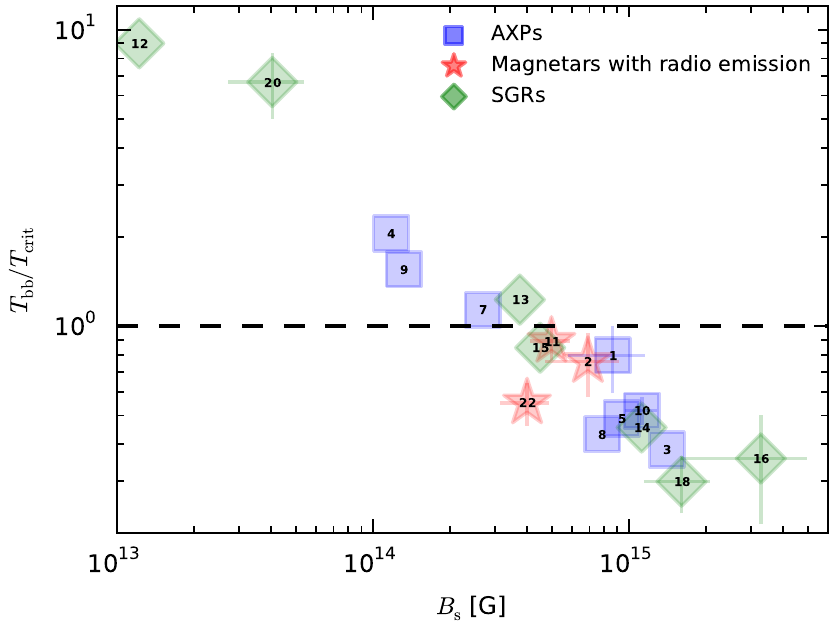}
        \end{center}
        \caption{
        Ratio of the observed surface/spot temperature to the predicted critical temperature of the polar cap. 
        The horizontal axis corresponds to the value of the inferred magnetic field at the polar cap.
        The red stars correspond to the magnetars with detected radio emission.
        }
        \label{fig:temp}
    \end{figure}
    
    It is widely believed that magnetars are very young neutron stars.
    As the approach to calculate the characteristic age, $\tau_{\rm c}$, assumes that the observed period is much longer than the rotational period at birth, $\tau_{\rm c}$ can be a poor approximation for their true age.
    Indeed, even a relatively low mean velocity of magnetars \citep{2013_Tendulkar} does not explain their very small Galactic scale height.
    Further evidence comes from estimates of ages of their host supernovae; thus, for example, the age of CTB 109 ($14 \, {\rm kyr}$) is much shorter than the characteristic age of the magnetar 1E 2259+586 ($\tau_{\rm c} = 230 \, {\rm kyr}$).
    As young neutron stars, magnetars tend to have high residual temperature.
    We mentioned in Section \ref{sec:psg} that a PSG can form if the surface temperature is lower than the critical value.
    Thus, we can define the first condition that must be met in order to allow formation of a PSG, and hence generation of radio emission, namely, the residual temperature has to be lower than the critical value $T_{\rm bb} < T_{\rm crit}$.
    In Figure \ref{fig:temp} we present the ratio of the observed residual temperature to the predicted critical temperature of the polar cap. 
    Confirming the hypothesis, all the magnetars with detected radio emission have a residual temperature below the critical value.
    Note that as a result of the fact that the residual temperature of SGR J1745-2900 is not known, it was not included in the figure.
    However, taking into account the upper limit for its X-ray luminosity in the quiescence mode, $L_{\rm bb}=1.1\times 10^{32} \, {\rm erg\,s^{-1}}$, the whole surface temperature of this magnetar is well below the critical value $T_{\rm bb}=6\times 10^{5} \, {\rm K} \ll 4.6\times 10^{6} \, {\rm K}$ ($T_{\rm bb}$ was calculated assuming $R_{\rm bb}=10\,{\rm km}$).
    Taking into account the temperature condition, we find that as long as the surface temperature will not decrease below the critical value, the following magnetars cannot generate radio emission: 1E 2259+586, 4U 0142+61, CXOU J164710.2-455216, SGR 0418+5729, SGR 0501+4516, Swift J1822.3-1606.

    \subsection{Power} \label{sec:strength}
    
    \begin{figure}[t]
        \begin{center}
            \includegraphics{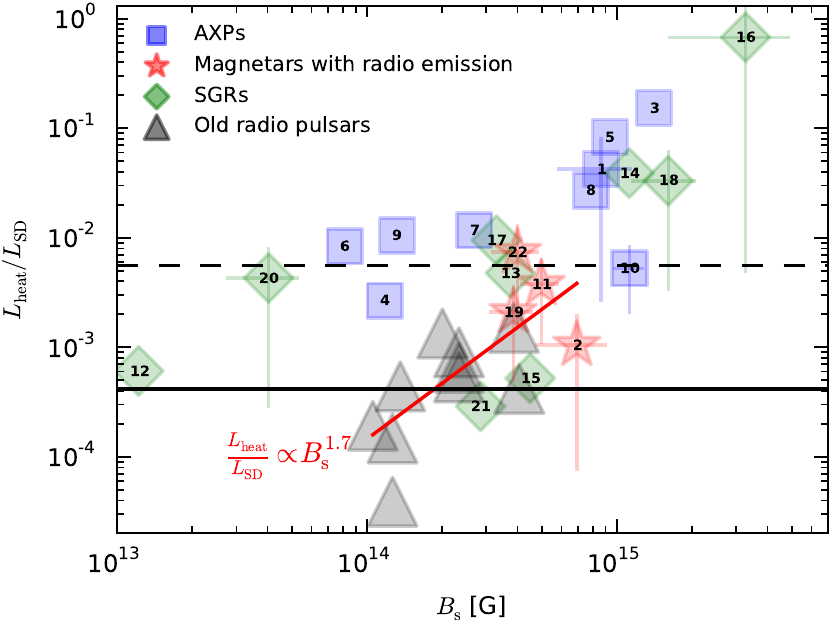}
        \end{center}
        \caption{
        Ratio of power required to heat up the polar cap to the critical temperature to the spin-down luminosity.
        The gray triangles show the observed thermal X-ray efficiency of radio pulsars. 
        The red stars correspond to magnetars with detected radio emission.
        The solid and dashed black horizontal lines correspond to the median and maximum values of the observed X-ray efficiency of radio pulsars, respectively.
        The red solid line corresponds to the linear fit for sources with radio emission (both magnetars and old radio pulsars).}
        \label{fig:efficiency}
    \end{figure}
    
    The high-energy radiation of magnetars (X-rays and $\gamma$-rays) during the active state is powered by the magnetic energy.
    Since the radio emission of magnetars appears only after the X-ray outburst it was believed that the magnetic energy is also a source of energy of the radio emission.
    However, based on the observed sample of radio magnetars, \cite{2011_Melikidze,2012_Rea} suggested that the radio emission from magnetars might be powered by the rotational energy.
    In normal radio pulsar particles are accelerated in the inner acceleration region at the expense of rotational energy giving the raise to plasma responsible for both radio emission and pulsar wind.
    If the plasma responsible for radio emission of magnetars is produced and accelerated in PSG (as in the case of radio pulsars) it should also be powered by the rotational energy.
    
    The strength of magnetic field at the polar cap of magnetars is of the order of $\sim 10^{14}-10^{15}\,{\rm G}$.
    Knowing the critical temperature (see Equation (\ref{t_crit})) and size of the polar cap, $A_{\rm pc}$, we can estimate power required to heat up the polar cap to the critical temperature:
    \begin{equation}
    	L_{\rm heat} = \sigma A_{\rm pc} T_{\rm crit}^4,
        \label{l_heat}
    \end{equation}
    where $\sigma$ is the Stefan-Boltzmann constant.
    In Figure \ref{fig:efficiency} we plot the ratio of power required to heat up the polar cap to the critical temperature to the spin-down luminosity. 
    For magnetars with detected radio emission, less then about $1\%$ of the rotational energy is enough to heat up the polar cap. 
    In the figure we present also the efficiency of X-ray emission of the polar cap for radio pulsars, $L_{\rm pc} / L_{\rm SD}$.
    Note that an estimation of an actual surface magnetic field of a pulsar is possible only for old sources for which radiation of the whole surface and the nonthermal magnetospheric X-ray emission do not dominate the X-ray spectrum.
    Similarly to radio magnetars, all radio pulsars are characterized by the polar cap radiation with luminosity considerably smaller than the spin-down luminosity $L_{\rm pc} / L_{\rm SD} < 1\%$.
    We argue that a magnetar can generate radio waves only if its spin-down luminosity is high enough, i.e., $L_{\rm heat} / L_{\rm SD} \lesssim 1\%$.
    Moreover, the power-law fit for all sources active in radio results in the following relationship:
    \begin{equation}
        L_{\rm pc,heat} / L_{\rm SD} \propto B_{\rm s}^{1.7}.
    \end{equation}
    It clearly shows that with increasing strength of the surface magnetic field at the polar cap, neutron stars active in radio (both pulsars and magnetars) use a greater part of their rotational kinetic energy to heat up the polar cap to the critical temperature and thus to form the PSG.
    Taking into account the power condition, we find that the radio emission of the following magnetars will be possible only after a significant decay of the magnetic field: 1E 1841-045, 1RXS J170849.0-400910, CXOU J010043.1-721134, and SGR 0526-66.
    Furthermore, taking into account the observational uncertainties the radio emission is unlikely to appear from the following sources: 1E 1048.1-5937, SGR 1806-20, and SGR 1900+14.

    \subsection{Visibility}
    
    \begin{figure}[t]
        \begin{center}
            \includegraphics{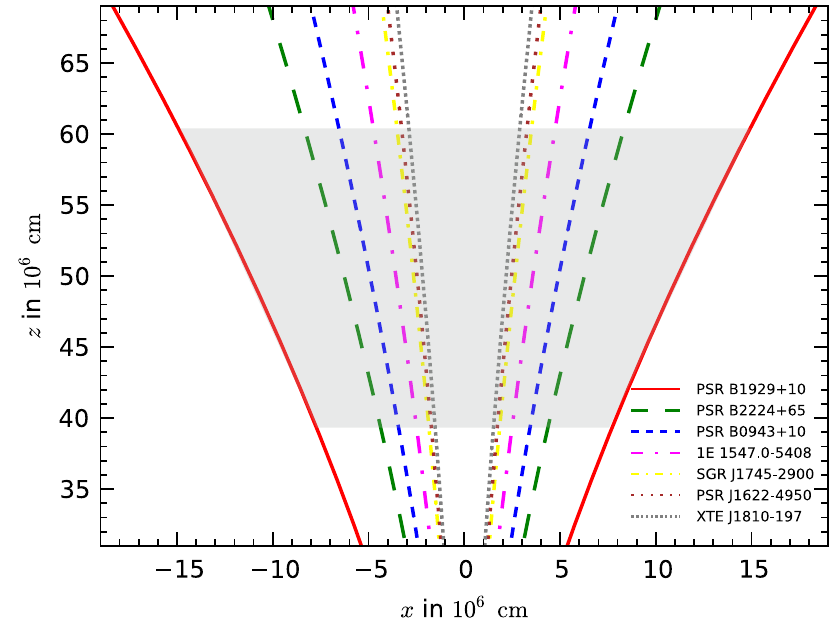}
        \end{center}
        \caption{
        Comparison of open field line regions of radio magnetars and a sample of radio pulsars. 
        The lines correspond to the last open magnetic field line. 
        }
        \label{fig:lines}
    \end{figure}
    
    In Sections \ref{sec:temperature} and \ref{sec:strength} we have shown the conditions that can be used to determine whether a magnetar can be active in radio.
    They are based on an assumption that plasma responsible for radio emission of magnetars is produced in the same manner as in radio pulsars (i.e., pair creation and acceleration in the PSG).
    However, we know from observations that in some aspects properties of magnetar radio emission are different from the properties of pulsar radio emission.
    Firstly, in magnetars radio emission is not continuous and appears only after the X-ray outburst.
    Secondly, the radio pulse profiles and fluxes vary on a timescales from minutes to days. 
    Finally, the average radio flux decays with decaying X-ray flux of the outburst.
    The bundle of open field lines of magnetars is much smaller than the one of normal pulsars.
    \cite{2009_Beloborodov} suggested that it may result in a narrow radio beam, and thus there is a small probability of it passing through our line of sight.
    As a consequence, when a magnetar is in quiescence, the radio pulsations are hardly detectable.
    The situation changes during the outburst when the magnetosphere is twisted.
    The magnetospheric currents alter the magnetic field in the region where radio emission is generated, thereby broadening the radio beam. 
    As the magnetosphere untwists, the radio beam returns to its original small size.
    Using the simple model (see, e.g., \cite{2004_Lorimer}), we can write that the opening angle of the radio beam is 
    \begin{equation}
        \rho \approx 0.4^{\circ}  R_{\rm em}^{0.5} P^{-0.5},
    \end{equation}    
    where $R_{\rm em}$ is the emission height in kilometers.
    Although a consensus regarding the radio emission process itself has not been established yet, the emission height is one of its least problematic aspects. 
    The PSG model, similarly as the vacuum gap model \citep{1975_Ruderman}, is the nonstationary model of the inner acceleration region.
    In such a model, the sparking-like gap discharges lead to the creation of pair plasma clouds.
    The plasma clouds consist of particles with a large spectrum of energies and move along the magnetic field lines.
    Particles from successive clouds overlap with each other owing to different energies, i.e. high-energy particles from a later cloud overlap with lower-energy particles from the earlier cloud.
    As shown by \cite{1998_Asseo}, this will result in an efficient two-stream instability that triggers electrostatic Langmuir waves.
    The electrostatic oscillations are unstable, which results in the formation of plasma solitons with a characteristic length along magnetic field lines of about $30 \,{\rm cm}$, thus making them capable of emitting coherent curvature radiation at radio wavelengths \citep{2000_Melikidze}.
    In the described model, the emission height is defined as a place where particles from two consecutive clouds overlap, leading to a two-stream plasma instability.
    The time after which particles with different Lorentz factors will overcome each other can be estimated as $t_{\rm r} \sim h / (2\Delta v)$, where $h$ is the gap height and $\Delta v\sim c/(2\gamma_{\rm p}^2)$ is the velocity difference, with $\gamma_{\rm p}$ being the average Lorentz factor of secondary plasma.
    Note the factor of two difference in the formula from the one presented in \cite{2000_Melikidze}.
    The difference is due to the fact that the vacuum gap model assumes that the plasma clouds are of the same size as the gap height, while in the PSG model the plasma clouds are considerably smaller than $h$.
    Furthermore, the vacuum gap model predicts the dependence of the gap height on pulsar period, dipolar component of the magnetic field, and curvature radius in the gap region.
    However, our latest studies \citep{2014_Szary} have shown that the radio luminosity does not depend on rotational parameters $P$ and $\dot{P}$.
    The result suggests that plasma responsible for radio emission is created and accelerated in similar conditions regardless of the dipolar component of the magnetic field.
    Indeed, in \cite{2015_Szary} we show that for wide ranges of magnetic field strength at the surface and curvature radius the gap height does not vary essentially and is of the order of $h\sim 50\, {\rm m}$.
    Finally, using the typical value for the average Lorentz factor of generated plasma, $\gamma_{\rm p}\approx 10^2$, we can estimate the emission height as $R_{\rm em}\approx t_{\rm r} c \approx h \gamma_{\rm p}^2 \sim 50R$, where $R$ is the neutron star radius.
    Note that the emission height may vary from one pulsar (or magnetar) to another, but as we have shown above, there is no theoretical justification that in magnetars the emission height should be higher than in normal radio pulsars.
    In Figure \ref{fig:lines} we show the comparison of the last open magnetic field lines for a sample of radio pulsars and magnetars with detected radio emission.
    It clearly shows that, owing to their longer periods, magnetars are characterized by a smaller radio beam.
    During the outburst, the curvature of open magnetic field lines of magnetars can significantly change, resulting in a much larger opening angle of radio emission.
    Furthermore, in the quiescent state of a magnetar, the curvature of magnetic field lines in the radio emission region is smaller than the one of radio pulsars, which may be of importance for the radio emission process. 
    It is also worth noting that during the outburst the flux of X-ray background photons increases, thereby facilitating inverse Compton scattering in the acceleration region, and thus the gap breakdown.

  \section{Conclusions}
    
    There are ongoing large theoretical and observational efforts to find when and which magnetars will emit radio waves.
    It was argued that in principle any magnetar undergoing an outburst could be radio active.
    Moreover, it was believed that whatever the mechanism of radio emission is, it should be different from that of rotation-powered radio pulsars.
    In this paper we show that both above statements are not true. 
    Furthermore, we show that not only can the observed sample of radio magnetars be explained within the framework of the PSG model, but we can use its predictions to establish whether a newly discovered magnetar will generate pulsed radio emission or not.
    Note that predictions regarding radio activity or inactivity of magnetars were performed assuming that the magnetic field at the polar cap does not differ significantly from the purely dipolar solution. 
    However, especially for the low-field magnetars, it may not be the case.
    As recently shown by \cite{2014_Tiengo}, the energy of the proton cyclotron absorption line in the X-ray spectrum of SGR 0418+5729 implies a magnetic field ranging from $2\times 10^{14} \, {\rm G}$ to more than $10^{15}\,{\rm G}$.
    The existence of small-scale, strong, multipolar components in an active magnetar is yet another feature that makes boundaries between magnetars and radio pulsars fade.

    \acknowledgments
  
    This work is supported by National Science Centre Poland under grants 2011/03/N/ST9/00669 and DEC-2012/05/B/ST9/03924.
    The data used in the paper are taken from the McGill magnetar catalog \citep{2014_Olausen}\footnote{www.physics.mcgill.ca/$\sim$pulsar/magnetar/main.html} and the catalog of isolated neutron stars with clearly observed thermal emission in quiescence by \citep{2013_Vigano} \footnote{www.neutronstarcooling.info}.

\end{document}